\documentclass[prl%
,twocolumngrid%
,superscriptaddress%
,floats%
]{revtex4}
\usepackage{dcolumn}%

\begin{document}
\title{Optimization with Extremal Dynamics} 
\date{\today}
\author{Stefan Boettcher} 
\email{sboettc@emory.edu}
\affiliation{Physics Department, Emory University, Atlanta, Georgia 30322, USA} 
\author{Allon G.\ Percus} 
\email{percus@lanl.gov}
\affiliation{Computer \& Computational Sciences Division, Los Alamos
National Laboratory, Los Alamos, NM 87545, USA}

\begin{abstract} 
We explore a new general-purpose heuristic for finding high-quality
solutions to hard optimization problems. The method, called extremal
optimization, is inspired by self-organized criticality, a concept
introduced to describe emergent complexity in physical
systems. Extremal optimization successively replaces extremely
undesirable variables of a single sub-optimal solution with new, random
ones. Large fluctuations ensue, that efficiently explore many local
optima. With only one adjustable parameter, the heuristic's performance has proven
competitive with more elaborate methods, especially near phase
transitions which are believed to coincide with the hardest instances.  We
use extremal optimization to elucidate the phase transition in the
3-coloring problem, and we provide independent confirmation of
previously reported extrapolations for the ground-state energy of $\pm
J$ spin glasses in $d=3$ and $4$.  \hfil\break 
PACS number(s): 02.60.Pn, 05.65.+b, 75.10.Nr, 64.60.Cn.
\end{abstract} 

\maketitle Many natural systems have, without any centralized
organizing facility, developed into complex structures that optimize
their use of resources in sophisticated ways~\cite{Bakbook}.
Biological evolution has formed efficient and strongly interdependent
networks in which resources rarely go to waste.  Even the morphology
of inanimate landscapes exhibits patterns that seem to serve a
purpose, such as the efficient drainage of water~\cite{Somfai,rivers}.

Natural systems that exhibit self-organizing qualities often possess a
common feature: a large number of strongly coupled entities with
similar properties.  Hence, at some coarse level they permit a
statistical description.  An external resource (sunlight in the case
of evolution) drives the system which then takes its direction purely
by chance. Like descending water breaking through the weakest of all
barriers in its wake, biological species are coupled in a global
comparative process that persistently washes away the least fit. In
this process, unlikely but highly adapted structures surface
inadvertently. Optimal adaptation thus emerges naturally, from the
dynamics, simply through a selection {\em against\/} the extremely
``bad''.  In fact, this process prevents the inflexibility inevitable
in a controlled breeding of the ``good''.

Various models relying on extremal processes have been proposed to
explain the phenomenon of self-organization~\cite{PMB}. In particular,
the Bak-Sneppen model of biological evolution is based on this
principle~\cite{BS,BoPa1}. Assuming an unspecified
interdependency between species, it produces salient nontrivial
features of pale-ontological data such as broadly distributed
lifetimes of species, large extinction events, and punctuated
equilibrium.

In the Bak-Sneppen model, species are located on the sites of a
lattice, and have an associated ``fitness'' value between 0 and 1.  At
each time step, the one species with the smallest value (poorest
degree of adaptation) is selected for a random update, having its
fitness replaced by a new value drawn randomly from a flat
distribution on the interval $[0,1]$.  But the change in fitness of
one species impacts the fitness of interrelated species.  Therefore,
all of the species at neighboring lattice sites have their fitness
replaced with new random numbers as well. After a sufficient number of
steps, the system reaches a highly correlated state known as
self-organized criticality (SOC)~\cite{BTW}.  In that state, almost
all species have reached a fitness above a certain threshold.  These
species, however, possess {\em punctuated equilibrium\/}: only one's weakened
neighbor can undermine one's own fitness. This coevolutionary
activity gives rise to chain reactions called ``avalanches'', large
fluctuations that rearrange major parts of the system, potentially
making any configuration accessible.

Although coevolution does not have optimization as its exclusive goal,
it serves as a powerful paradigm.  We have used it as motivation for a
new approach to optimization~\cite{BoPe1}.  The heuristic we have
introduced, called extremal optimization (EO), follows the spirit of
the Bak-Sneppen model, merely updating those variables having
the ``worst'' values in a solution and replacing them by random values
without ever explicitly improving them.

Previously, several heuristics inspired by natural processes have been
proposed, notably {\em simulated annealing}~\cite{Science} and {\em
genetic algorithms}~\cite{Holland}.  These have aroused considerable
interest among physicists, in the development of physically motivated
heuristics~\cite{Houdayer,Dittes,Moebius}, and in their applications
to physical problems~\cite{Pal,Hartmann_d3,Hartmann_d4}.  EO adds a
distinctly new approach to the approximation of hard optimization
problems by utilizing large fluctuations inherent to systems driven far from
equilibrium. In this Letter we demonstrate EO's generality, the
simplicity of its implementation, and its results, for a physical
problem as well as for a classic combinatorial optimization problem. 
A tutorial introduction is given in Ref.~\cite{CISE}.

The physical problem we consider is a spin glass~\cite{MPV}.  It
consists of a $d$-dimensional hyper-cubic lattice of length $L$, with
a spin variable $x_i\in\{-1,1\}$ at each site $i$, $1\leq i\leq n$
$(=L^d)$.  A spin is connected to each of its nearest neighbors $j$
via a bond variable $J_{ij}\in\{-1,1\}$, assigned at random. The
configuration space $\Omega$ consist of all configurations
$S=(x_1,\ldots,x_n)\in\Omega$ where $|\Omega|=2^n$.

We wish to minimize the cost function, or Hamiltonian
\begin{eqnarray}
C(S)=H({\bf x})=-{1\over2}{\sum\sum}_{\langle i,j\rangle}J_{ij}x_ix_j.
\label{spineq}
\end{eqnarray}
Due to frustration~\cite{MPV}, ground state configurations $S_{\rm
min}$ are hard to find, and it has been shown that for $d>2$ the
problem is among the hardest optimization problems~\cite{Barahona}.

To find low-energy configurations, EO assigns a fitness to each spin
variable $x_i$
\begin{eqnarray}
\lambda_i=x_i\,\left({1\over2}\sum_j J_{ij}x_j\right),
\label{lambdaeq}
\end{eqnarray}
so that
\begin{eqnarray}
C(S)=-\sum_{i=1}^n \lambda_i.
\label{costeq}
\end{eqnarray}
Unlike in genetic algorithms, the fitness does not characterize an
entire configuration, but simply a single variable of the
configuration.  In similarity to the Bak-Sneppen model, EO then
proceeds to search $\Omega$ by sequentially changing variables with
``bad'' fitness at each update step. The simplest ``neighborhood'' $N(S)$
for an update consists of all configurations $S'\in N(S)$ that could
be reached from $S$ through the flip of a single spin. After each
update, the fitnesses of the changed variable and of all its neighbors
are reevaluated according to Eq.~(\ref{lambdaeq}).

The basic EO algorithm proceeds as follows:
\begin{center}
\framebox[3.3in]{
\begin{minipage}[t]{3.1in}
\begin{enumerate}
\item Initialize configuration $S$ at will; set $S_{\rm best}\!:=\!S$.
\item For the ``current'' configuration $S$,
\label{EOupdate}
\begin{enumerate}
\item evaluate $\lambda_i$ for each variable $x_i$,
\label{evaluate}
\item find $j$ satisfying $\lambda_j\leq\lambda_i$ for all $i$, {\em
i.e.,\/} $x_j$ has the ``worst fitness'',
\label{sort}
\item choose $S'\!\in\!N(S)$ so that $x_j$ {\em must} change,%
\label{worst}
\item accept $S:=S'$ {\em unconditionally,}
\label{alwaysmove}
\item if $C(S)<C(S_{\rm best})$ then set $S_{\rm best}:=S$.
\end{enumerate}
\item Repeat at step~(\ref{EOupdate}) as long as desired.
\item Return $S_{\rm best}$ and $C(S_{\rm best})$.
\end{enumerate}
\end{minipage}}
\end{center}

Initial tests have shown that this basic algorithm is quite
competitive for optimization problems where EO can choose randomly
among many $S'\in N(S)$ satisfying step~(\ref{worst}), such as graph
partitioning~\cite{BoPe1}.  But, in cases such as the single spin-flip
neighborhood above, focusing on only the worst fitness
[step~(\ref{sort})] leads to a deterministic process, leaving no
choice in step~(\ref{worst}). To avoid such ``dead ends'' and to
improve results~\cite{BoPe1}, we introduce a single parameter into the
algorithm. We rank all the variables $x_i$ according to fitness
$\lambda_i$, {\em i.e.\/}, find a permutation $\Pi$ of the variable
labels $i$ with
\begin{eqnarray}
\lambda_{\Pi(1)}\leq\lambda_{\Pi(2)}\leq\ldots\leq\lambda_{\Pi(n)}.
\end{eqnarray}
The worst variable $x_j$ [step~(\ref{sort})] is of rank 1, $j=\Pi(1)$,
and the best variable is of rank $n$. Now, consider a probability
distribution over the {\em ranks\/} $k$,
\begin{eqnarray}
P_k\propto k^{-\tau},\qquad 1\leq k\leq n,
\label{taueq}
\end{eqnarray}
for a given value of the parameter $\tau$. At each update, select a
rank $k$ according to $P_k$.  Then, modify step~(\ref{worst}) so that
the variable $x_i$ with $i=\Pi(k)$ changes its state.

\begin{figure}
\vskip 2.2in \includegraphics{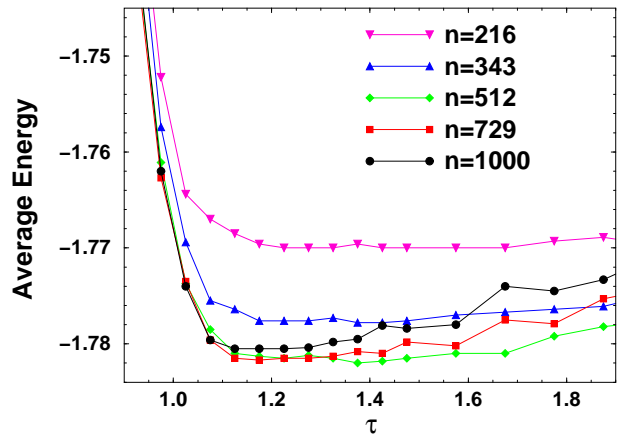}
\caption{Plot of the average energies obtained by EO for
the $\pm J$ spin glass in $d=3$ as a function of $\tau$.
For each size $n$, 10 instances were chosen.  For each instance, 10 EO 
runs were performed starting from different initial conditions at each
$\tau$.  The results were averaged over runs and over instances.
The best results are obtained at $\tau$ moving slowly toward
$\tau\to1^+$ for $n\to\infty$.
}
\label{glasstau}
\end{figure}

For $\tau=0$, this ``$\tau$-EO'' algorithm is simply a random walk
through $\Omega$.  Conversely, for $\tau\to\infty$, the process
approaches a deterministic local search, only updating the
lowest-ranked variable, and is bound to reach a dead end (see
Fig.~\ref{glasstau}). However, for finite values of $\tau$ the choice
of a scale-free distribution for $P_k$ in Eq.~(\ref{taueq}) ensures
that no rank gets excluded from further evolution, while 
maintaining a clear bias against variables with bad fitness.

For the spin glass, we obtained our best solutions for $\tau\to1^+$,
as shown in Fig.~\ref{glasstau}.  Generally, over many optimization
problems, the preferred value seems to scale slowly as
$\tau-1\sim1/\ln n$ for increasing $n$.  Experiments with random
graphs support this expectation~\cite{BoPe2}, although the dependence
on $n$ can be practically negligible. For instance, at fixed
$\tau=1.4$, $\tau$-EO already reproduced many testbed results for the
partitioning of graphs of sizes $n=1000$ to
$n\gtrsim10^5$~\cite{BoPe1}.

We have run the $\tau$-EO algorithm with $\tau=1.15$ on a large number
 of realizations of the $J_{ij}$, for $n=L^d$ with $L=5,6,7,8,9,10,12$
 in $d=3$, and with $L=3,4,5,6,7$ in $d=4$. To reduce variances, we
 fixed $|\sum J_{ij}| \leq1$. For each instance, we have run EO with 5
 restarts from random initial conditions, retaining only the lowest
 energy state obtained, and then averaging over instances. Inspection
 of the convergence results for the genetic algorithms  in
 Refs.~\cite{Pal,Houdayer} suggest a runtime scaling at least as
 $n^3$--$n^4$ for consistent performance. Indeed, using $\sim n^4/100$
 updates enables EO to reproduce its lowest energy states on about 80\% to
 95\% of the restarts, for each $n$. Our results are listed in
 Table~\ref{table1}. A fit of our data with $e_d(n)=e_d(\infty)+A/n$ for
 $n\to\infty$ predicts $e_3(\infty)=1.7865(3)$ for $d=3$ and
 $e_4(\infty)=2.093(1)$ for $d=4$. Both values are consistent with
 the findings of Refs.~\cite{Pal,Hartmann_d3,Hartmann_d4}, providing
 independent confirmation of those results with far less parameter
 tuning.

\begin{table*}
\caption{EO approximations to the average ground-state energy per spin
$e_d(n)$ of the $\pm J$ spin glass in $d=3$ on the left, compared with
genetic algorithm results from Refs.~\protect\cite{Pal,Hartmann_d3}, and in $d=4$ on
the right (see also Ref.~\cite{Hartmann_d4}). For each size $n=L^d$ we
have studied a large number $I$ of instances. Also shown is the
average time $t$ (in seconds) needed for EO to find the presumed
ground state, on a 450MHz Pentium.}
\begin{tabular}{r|rldll|rld}
\toprule
$L$ & $I$ & $e_3(n)$ & $t$ & Ref.~\protect\cite{Pal} &
Ref.~\protect\cite{Hartmann_d3} & $I$ & $e_4(n)$ & $t$\\
\hline
     3  &40100  & -1.6712(6) &0.0006  &-1.67171(9) & -1.6731(19) &  10000 &  -2.0214(6)& 0.0164\\
     4  &40100  & -1.7377(3) &0.0071  &-1.73749(8) &-1.7370(9) &   4472 &  -2.0701(4)& 0.452\\
    5  &28354  & -1.7609(2) &0.0653  &-1.76090(12)&-1.7603(8) &   2886 &  -2.0836(3)& 8.09\\
    6  &12937  & -1.7712(2) &0.524  &-1.77130(12)& -1.7723(7) &  283 &  -2.0886(6)& 86.3\\
    7  & 5936  & -1.7764(3) &3.87  & -1.77706(17) &  &    32 &  -2.0909(12)& 1090.\\
    8  & 1380  & -1.7796(5) &22.1  &-1.77991(22)&-1.7802(5)&&&\\
    9  &  837  & -1.7822(5) &100. &&&&&\\
   10  &  777  & -1.7832(5) &424.  &-1.78339(27)  &-1.7840(4)&&&\\ 
   12  &   30  & -1.7857(16) &9720.  &-1.78407(121) &-1.7851(4)&&&\\
\botrule
\end{tabular}
\label{table1}
\end{table*}

In the future, we intend to use EO to explore the properties of states
near the ground state. As the results on 3-coloring in Ref.~\cite{BGIP}
suggest, EO's continued fluctuations through
near-optimal configurations even after first reaching near-optimal states
may provide an efficient means of exploring a configuration space widely.

To demonstrate the versatility of EO (see also Ref.~\cite{PPSN}), we
now turn to a popular combinatorial optimization problem, graph
coloring~\cite{G+J}. Specifically, we consider random graph
3-coloring~\cite{AI,Culberson}. A random graph is generated by
connecting any pair of its $n$ vertices by an edge, with probability
$p$~\cite{Bollobas}.  In $K$-coloring, given $K$ different colors used
to label the vertices of a graph, we need to find a coloring that
minimizes the number of ``monochromatic'' edges connecting vertices of
identical color.  We implement EO by defining $\lambda_i$ for each
vertex to be $-1/2$ times the number of monochromatic edges attached to
it.  Then, Eq.~(\ref{costeq}) represents exactly the cost function, counting 
the number of monochromatic edges present.  As a simple neighborhood 
definition, at each update we merely change the color of one ``bad'' 
vertex selected according to $\tau$-EO [step~(\ref{worst})].

As a special challenge, we have used $\tau$-EO for 3-coloring to
investigate the phase transition that occurs under variation of the
average vertex degree $\alpha=pn$~\cite{Cheeseman,Culberson}.  Random
graphs with small $\alpha$ can almost always be colored at zero cost,
while graphs with large $\alpha$ are typically very homogeneous with a
high but easily approximated cost. Located at some point $\alpha_{\rm
crit}$ between these extremes, there is a sharp phase transition to
non-zero cost solutions. Such a critical point appears in many
combinatorial optimization problems, and has been conjectured to
harbor those instances that are the hardest to solve
computationally~\cite{Cheeseman}.  Previously we have
shown~\cite{EOperc} that $\tau$-EO significantly outperforms simulated
annealing near the phase transition of the bipartitioning problem of
random graphs.

Using EO we can estimate the value of $\alpha_{\rm crit}$ for
3-coloring. To this end, we have averaged the cost EO obtains as a
function of the vertex degree $\alpha$.  We generated $10000$, $5000$,
$1300$, and $650$ instances for $n=32$, 64, 128, and 256,
respectively, for values of $3.6\leq\alpha\leq6$. Since $n$ is
relatively small and the runs were chosen to be very long ($100n^2$ updates),
we found optimal performance at $\tau=2.7$.  Such excessively long
runs were used as part of a study to find {\em all\/} minimal-cost
solutions for each instance, in order to determine their overlap (or
``backbone''). Elsewhere~\cite{BGIP} we show that this backbone
appears to undergo a first-order phase transition as conjectured in
Ref.~\cite{Monasson}.

\begin{figure}
\vskip 2.2in  
\includegraphics{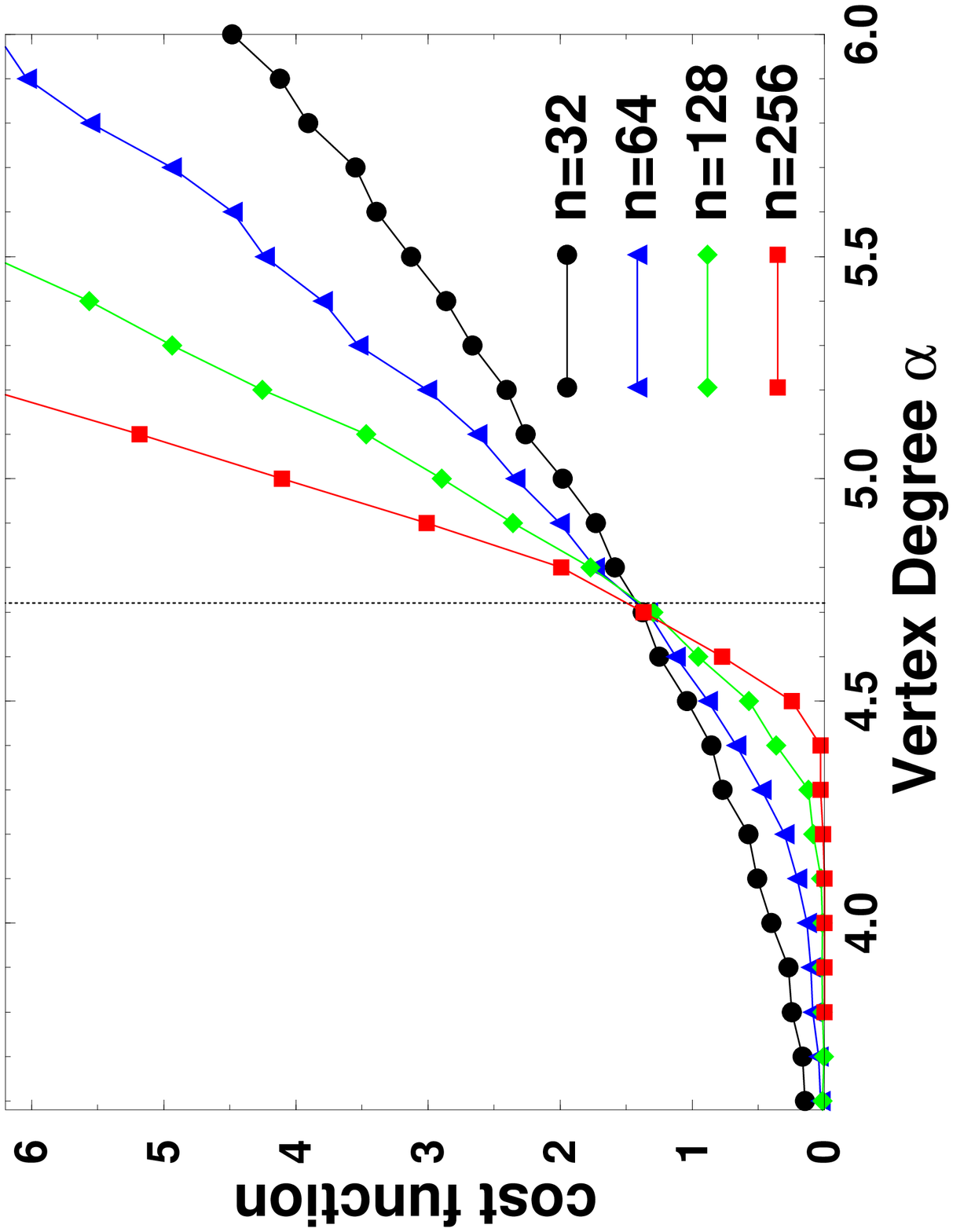}   
\includegraphics{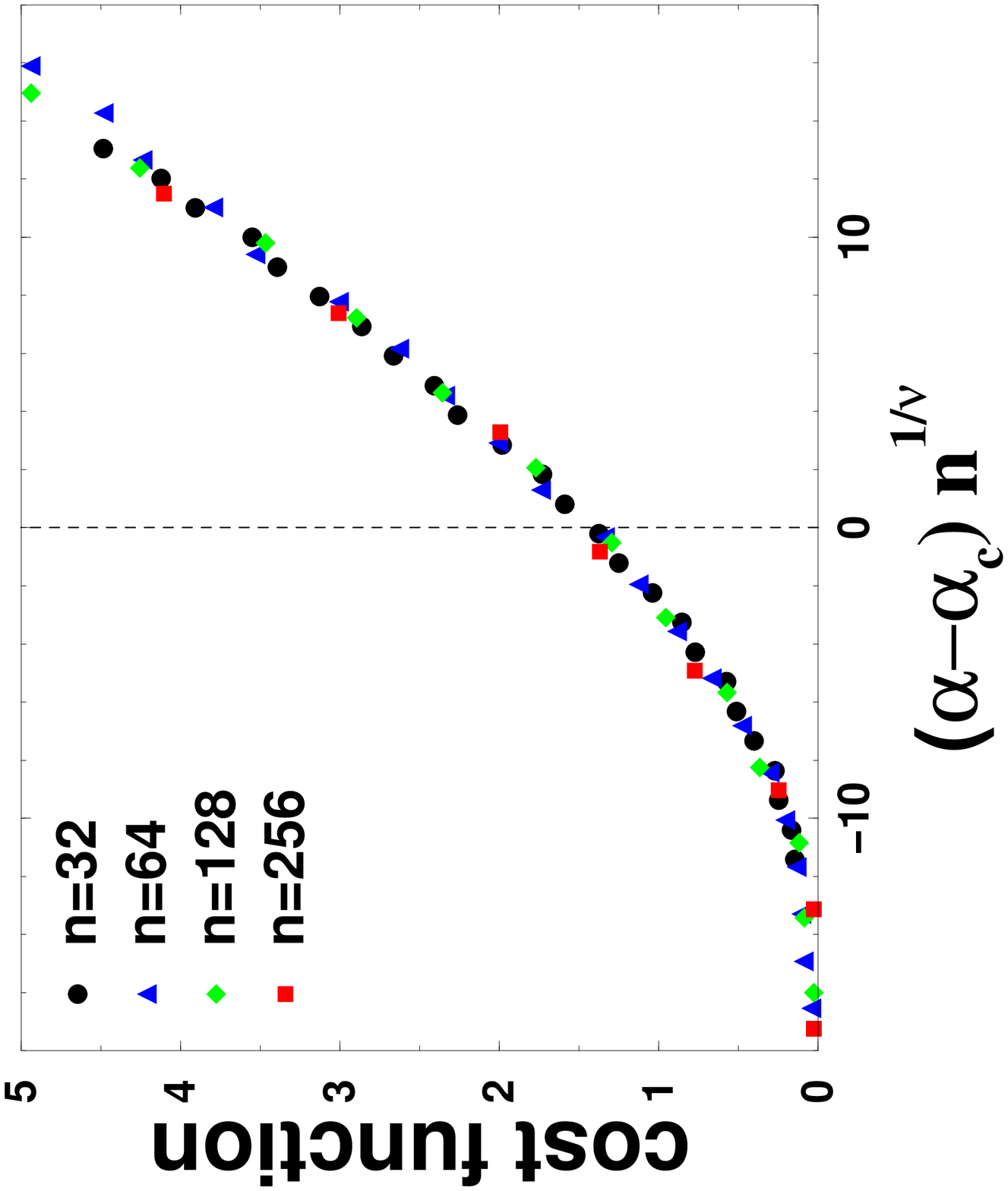}
\caption{Plot of the average cost as a function of the average vertex
degree $\alpha$ for random graph 3-coloring. According to
Eq.~(\protect\ref{scaleq}), the data collapse in the insert predicts
$\alpha_{\rm crit}\approx4.72$, indicated by a vertical line.}
\label{3colcost}
\end{figure}

Finite size scaling with the ansatz
\begin{eqnarray}
\langle C\rangle(\alpha,n)\sim f\left[\left(\alpha-\alpha_{\rm
crit}\right) n^{1/\nu}\right]
\label{scaleq}
\end{eqnarray}
applied to the results depicted in Fig.~\ref{3colcost}  predicts
$\alpha_{\rm crit}\approx4.72(1)$ and $\nu\approx1.53(5)$.  These  are
the most precise estimates to date.  The numerical value of $\nu$
suggests that in fact $\nu=3/2$, just as for the percolation
transition of random graphs~\cite{Bollobas} and the transition in
3-satisfiability~\cite{KS}.

In conclusion, we have presented a new optimization method, called
extremal optimization due to its derivation from extremally driven
statistical systems.  At each update step, the algorithm assigns
fitnesses to variables $x_i$, and then generates moves by randomly
updating an ``unfit'' variable.  EO gives no consideration to the
move's outcome.  Large fluctuations in the cost function can accumulate
over many updates; only the bias {\em against\/} poor fitnesses guides
EO back towards improved solutions.

A drawback to EO is that a general definition of fitness for
individual variables may prove ambiguous or even impossible.  Also, each
variable $x_i$ could have a large number of states to choose from, as in
$K$-coloring with large $K$; random updates would then be more likely to
remove than to create well-adapted variables (this is notably the case for
the traveling
salesman problem~\cite{BoPe1}). And in highly connected systems (e. g. for $\alpha\gg1$ in $K$-coloring), EO may be
slowed down considerably by continual fitness recalculations
[step~(\ref{evaluate})].

However, extremal optimization is readily applicable to problems whose
cost can be decomposed into contributions from individual degrees of
freedom.  It is easily implemented and, using very few parameters, it
can prove highly competitive.  We have shown this on the spin glass
Hamiltonian, obtaining $d=3$ and $4$ ground-state energies that are
consistent with the best known results.  We have also used EO to
explore the phase transition in random graph 3-coloring.  Its results
enable us to provide, by way of finite size scaling, the first sound
estimates of critical values for this problem.

\begin{acknowledgments}
We would like to thank the participants of the 1999 Telluride Workshop
on Energy Landscapes for valuable input.  Special thanks to Paolo
Sibani, Jesper Dall, Sigismund Kobe, and Gabriel Istrate.  This work
was supported in part by the URC at Emory University and by an LDRD grant
from Los Alamos National Laboratory.
\end{acknowledgments}

\end{document}